# Extremely Strong Coupling Superconductivity in Artificial Two-dimensional Kondo Lattices

Y. Mizukami[1], H. Shishido[1,+], T. Shibauchi[1], M. Shimozawa[1], S. Yasumoto[1], D. Watanabe[1], M. Yamashita[1], H. Ikeda[1], T. Terashima[2], H. Kontani[3], and Y. Matsuda[1,*]

[1] *Department of Physics, Kyoto University, Kyoto 606-8502, Japan*
[2] *Research Center for Low Temperature and Materials Sciences, Kyoto University, Kyoto 606-8501, Japan*
[3] *Department of Physics, Nagoya University, Furo-cho, Nagoya 464-8602, Japan*
+*Present address: Department of Physics and Electronics, Osaka Prefecture University, Sakai, Osaka 599-8531, Japan*
*matsuda@scphys.kyoto-u.ac.jp

**When interacting electrons are confined to low-dimensions, the electron-electron correlation effect is enhanced dramatically, which often drives the system into exhibiting behaviours that are otherwise highly improbable. Superconductivity with the strongest electron correlations is achieved in heavy-fermion compounds, which contain a dense lattice of localized magnetic moments interacting with a sea of conduction electrons to form a three-dimensional (3D) Kondo lattice[1]. It had remained an unanswered question whether superconductivity would persist upon effectively reducing the dimensionality of these materials from three to two. Here we report on the observation of superconductivity in such an ultimately strongly-correlated system of heavy electrons confined within a 2D square-lattice of Ce-atoms (2D Kondo lattice), which was realized by fabricating epitaxial superlattices[2,3] built of alternating layers of heavy-fermion $CeCoIn_5$[4] and conventional metal $YbCoIn_5$. The field-temperature phase diagram of the superlattices exhibits highly unusual behaviours, including a striking enhancement of the upper critical field relative to the transition temperature. This implies that the force holding together the superconducting electron-pairs takes on an extremely strong coupled nature as a result of two-dimensionalisation.**

The layered heavy-fermion compound $CeCoIn_5$ has the highest superconducting transition temperature ($T_c$=2.3 K) among rare-earth-based heavy-fermion materials[4]. Its electronic properties are characterized by anomalously large value of the linear contribution to the specific heat (Sommerfeld coefficient $\gamma \sim 1$ J/mol K$^2$) indicating heavy effective masses of the $4f$ electrons which contribute greatly to the Fermi surface. The tetragonal $CeCoIn_5$ crystal structure is built from alternating layers of $CeIn_3$ and $CoIn_2$ stacked along the [001] direction. This compound possesses several key features for understanding the unconventional superconductivity in strongly correlated systems[5-7]. The superconductivity with $d_{x^2-y^2}$ pairing symmetry[8-11] which occurs in the proximity of a magnetic instability is a manifestation of magnetic fluctuations mediated superconductivity[5-7,12]. A very strong coupling superconductivity, where electron-pairs are bound together by strong forces, is revealed by a large specific heat jump[4] at $T_c$ representing a steep drop of the entropy below $T_c$, and a large superconducting energy gap $\Delta$ needed to break the electron-pair[9]. Despite its layered structure, the largely corrugated Fermi surface[13], 3D-like antiferromagnetic fluctuations in the normal state[14], and small anisotropy of upper critical field[15], all indicate that the electronic, magnetic and superconducting properties are essentially 3D rather than 2D. Therefore it is still unclear to which extent the 3D nature is essential for the superconductivity of $CeCoIn_5$.

Recently the state-of-the-art technique has been developed to reduce the dimensionality of the heavy electrons in a controllable fashion by the layer-by-layer epitaxial growth of Ce-based materials. Previously a series of antiferromagnetic superlattices $CeIn_3/LaIn_3$ have been successfully grown[2], but it remains open whether heavy electrons in a single Ce-layer forming a 2D Kondo lattice can be superconducting. Here we fabricate multilayers of $CeCoIn_5$ sandwiched by nonmagnetic and nonsuperconducting metal $YbCoIn_5$ (Yb-ion is divalent in closed-shell $4f(14)$ configuration) forming ($n$:$m$) c-axis oriented superlattice structure, where $n$ and $m$ are the number of layers of $CeCoIn_5$ and $YbCoIn_5$ in a unit cell, respectively. Small lattice mismatch between $CeCoIn_5$ and $YbCoIn_5$ offers a possibility of providing an ideal heterostructure. The high resolution cross-sectional transmission-electron-microscope (TEM) results (Figs. 1a-c), and distinct lateral satellite peaks in X-ray diffraction pattern (Fig. 1d) demonstrate the continuous and evenly spaced $CeCoIn_5$ layers with no discernible interdiffusion even for $n$=1 cases (see Supplementary Fig. S1 for quantitative analysis of interdiffusion by X-ray). The epitaxial growth of each layer with atomic flatness is shown by the streak patterns of the reflection-high-energy-electron-diffraction (RHEED) (Fig.1e) and atomic-force-microscopy (AFM) image (Fig. 1f). We investigate the transport properties for the ($n$:5) superlattices by varying $n$. The resistivity of $CeCoIn_5$ thin film (Fig. 2a) reproduces well that of bulk single crystals[4]; Below ~100 K $\rho(T)$ increases upon cooling due to the Kondo scattering, decreases after showing a peak at around the coherence temperature $T_{coh}$~30 K, and drops to zero at the superconducting transition. The hump structure of $\rho(T)$ at ~$T_{coh}$ is also observed in the superlattices but becomes less pronounced with decreasing $n$. The superconducting transition to zero resistance is observed in the superlattices for *n*≥3 (Fig. 2b). For *n*=2 and 1, ρ(*T*) decreases below ~1 K, but it does not reach zero. However, when the magnetic field is applied perpendicular to the layers for *n*=1, ρ(*T*) increases and recovers to the value extrapolated above 1 K at 5 T, while the reduction of ρ(*T*) below 1 K remains in the parallel field of 6 T (Fig. 2c). The observed large and anisotropic field response of ρ(*T*) is typical for layered superconductors, demonstrating the superconductivity even in *n*=1 superlattice with 2D square lattice of Ce-atoms. The critical temperature $T_c$ determined by the resistive transition gradually decreases with decreasing *n* (Fig. 2d). The residual resistivity $\rho_0$ of the superlattices is in the same order as $\rho_0$ of single-crystalline film (Fig. 2d) and is much lower than $\rho_0$ of Yb-substituted CeCoIn$_5$ single crystals[16]. An important question is whether the superconducting electrons in the superlattices are heavy and if so what is their dimensionality. As displayed in Figs. 2c, 3a and 4a, the parallel and perpendicular (to the layers) upper critical field, $H_{c2\parallel}$ and $H_{c2\perp}$, of the superlattices at low temperature are significantly larger than those in conventional superconductors with similar $T_c$. The magnetic field destroys the superconductivity in two distinct ways, the orbital pair-breaking effect (vortex formation) and Pauli paramagnetic effect, a breaking up of pair by spin-polarization. The zero-temperature value of the orbital upper critical field in perpendicular field $H_{c2\perp}^{\mathrm{orb}}(0)$ reflects the effective electron mass in the plane $m_{ab}^*$, $H_{c2\perp}^{\mathrm{orb}}(0) \propto m_{ab}^{*2}$, and is estimated to be 6, 11, 12 T for *n*=3, 5, and 7 superlattices from the initial slope of $H_{c2\perp}$ at $T_c$ by the relation, $H_{c2\perp}^{\mathrm{orb}}(0)=0.69T_c(-dH_{c2\perp}/dT)_{T_c}$. These magnitudes are comparable or in the same order of $H_{c2\perp}^{\mathrm{orb}}(0)$ (=14 T) in bulk single crystal, providing strong evidence for the superconducting ``heavy'' electrons in the superlattices. We stress that even a slight deviation of the *f*-electron number from unity leads to a serious reduction of the heavy electron mass[17]. Moreover the band structure calculation for $n$=1 superlattice shows the number of $f$-electrons in each CeCoIn$_5$ layer is very close to unity (Supplementary Information). These indicate that the $f$-electron wave functions are essentially confined to Ce-layers. The magnetic two-dimensionality is shown by estimating the strength of Ruderman-Kittel-Kasuya-Yosida (RKKY) interaction, an intersite magnetic exchange interaction between the localized $f$-moments, which decays with the distance as $1/r^3$. This interaction between the Ce-ions in different layers of (n:5) superlattices reduces to less than 1/200 of that between neighboring Ce-ions within the same layer.

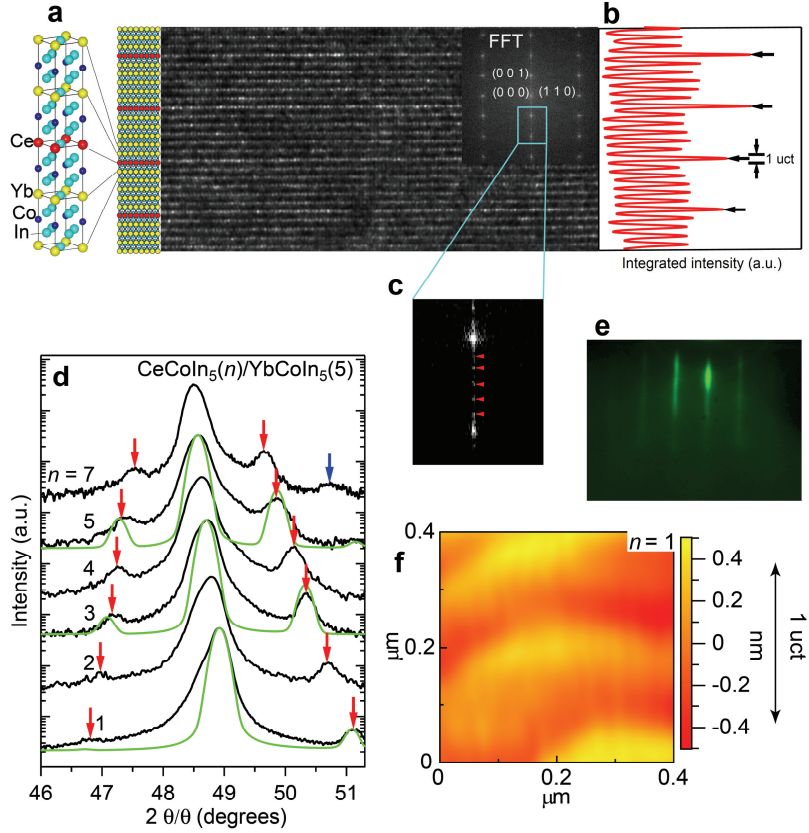

**Figure 1 | Epitaxial superlattices (*n*:5) of CeCoIn$_5$(n)/YbCoIn$_5$(5). a**, High-resolution cross sectional TEM image of *n*=1 superlattice. The bright dot arrays are identified as the Ce layers and the less bright dots are Yb atoms, which is consistent with the designed superlattice structure in the left panel. The intensity integrated over the horizontal width of the image plotted against vertical position (**b**) indicates clear difference between the Ce and Yb layers, showing no discernible atomic interdiffusion between the neighboring Ce and Yb layers. Upper right inset is the fast Fourier transform (FFT) of the TEM image, which shows clear superspots along the [001] direction (**c**). **d**, Cu-K$_{\alpha1}$ x-ray diffraction patterns for *n*=1, 2, 3, 4, 5, and 7 superlattices with typical total thickness of 300 nm show first (red arrows) and second (blue arrow) satellite peaks. The position of the satellite peaks and their asymmetric heights can be reproduced by the step-model simulations (green lines) neglecting interface and layer-thickness fluctuations[29] (see also Supplementary Information for the detailed analysis of the satellite peak intensity). **e**, Streak patterns of the RHEED image during the growth. **f**, Typical AFM image for *n*=1 superlattice. Typical surface roughness is within 0.8 nm, comparable to one unit-cell-thickness along the *c* axis of CeCoIn$_5$.

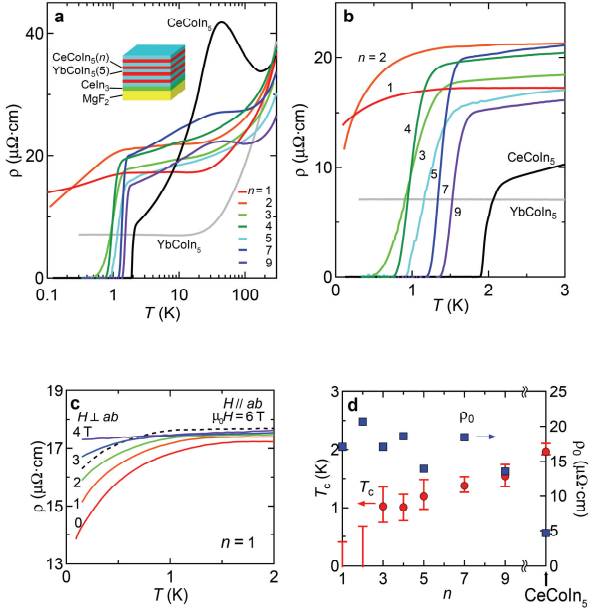

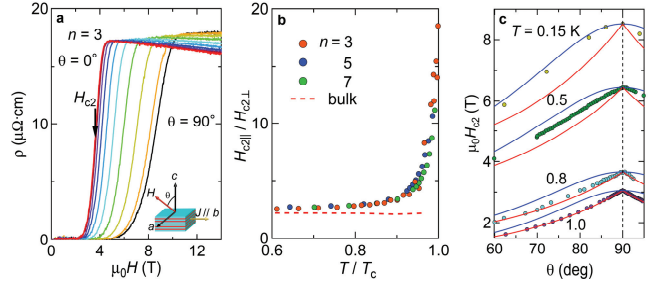

**Figure 2 | Superconductivity in superlattices (n:5) of CeCoIn$_5$(n)/YbCoIn$_5$(5). a**, Temperature dependence of electrical resistivity $\rho(T)$ for $n$=1, 2, 3, 5, 7, and 9, compared with those of 300-nm-thick CeCoIn$_5$ and YbCoIn$_5$ single-crystalline thin films. **b**, Low-temperature part of the same data as in **a**. **c**, $\rho(T)$ for $n$=1 at low temperatures in magnetic field parallel (dotted line) and perpendicular (solid lines) to the $ab$ plane. **d**, Superconducting transition temperature as a function of $n$ (left axis). The circles are the mid points of the resistive transition and the bars indicate the onset and zero-resistivity temperatures. The residual resistivity $\rho_0$ as a function of $n$ is also shown (right axis).

**Figure 3 | Superconducting anisotropy in superlattices (n:5) of CeCoIn$_5$(n)/YbCoIn$_5$(5). a**, Magnetic-field dependence of the resistivity for $n$=3 superlattice at several field angles from θ=0 (***H***⊥$ab$) to 90 deg (***H***//$c$) (10 deg step) at $T$=0.1 K. **b**, Anisotropy of $H_{c2}$, $H_{c2\parallel}/H_{c2\perp}$, as a function of reduced temperature $T/T_c$ for $n$=3, 5, and 7 superlattices and for the bulk CeCoIn$_5$. **c**, Upper critical field $H_{c2}(\theta)$ at several temperatures as a function of field angle θ. $H_{c2}$ is determined by the mid-point of the transition except for 1.0 K, where a 80% resistivity criterion has been used. The solid blue and red lines are the fits to the 3D anisotropic mass model represented as $H_{c2}(\theta) = H_{c2\parallel}/(\sin^2\theta+\gamma^2\cos^2\theta)^{1/2}$ with $\gamma = H_{c2\parallel}/H_{c2\perp}$ and Tinkham's formula $|H_{c2}(\theta)\cos\theta/H_{c2\perp}|+\{H_{c2}(\theta)\sin\theta/H_{c2\parallel}\}^2=1$ for a 2D superconductor, respectively[19].

The superconducting order parameters in the CeCoIn$_5$ layers of the superlattices are expected to be coupled weakly by the proximity effect through the normal metal YbCoIn$_5$ layers. The proximity induced superconductivity in YbCoIn$_5$ layers is expected to be very fragile and destroyed when a weak field is applied[18]. If the thickness of CeCoIn$_5$ layer is comparable to the perpendicular coherence length $\xi_\perp$ (~2.1 nm for CeCoIn$_5$) and the separation of superconducting layers (~3.7 nm for (n:5) superlattices) exceeds $\xi_\perp$, each CeCoIn$_5$ layer acts as a 2D superconductor[19]. This 2D feature is revealed by diverging $H_{c2\parallel}/H_{c2\perp}$ of $n$=3, 5 and 7 superlattices with approaching $T_c$ (Fig. 3b) in sharp contrast to bulk CeCoIn$_5$[20], and by a cusp-like angular dependence $H_{c2}(\theta)$ near parallel field for $n$=3 superlattice (Fig. 3c), which is qualitatively different from that expected in the 3D anisotropic-mass-model but is well fitted by the model in 2D limit[19]. Based on the above 2D features observed in all electronic, magnetic and superconducting properties, we conclude that the observed heavy-electron superconductivity is mediated most likely by 2D electron correlation effects. A fascinating issue is how the two-dimensionalisation changes the pairing nature. The fact that $H_{c2\perp}^{\mathrm{orb}}(0)$ estimated from the initial slope of $H_{c2\perp}(T)$ at $T_c$ well exceeds the actual $H_{c2\perp}$ at low temperatures indicates the predominant Pauli paramagnetic pair-breaking effect even in perpendicular field. Therefore $H_{c2}(\theta)$ at low temperatures is dominated by the Pauli effect in any field directions. This is reinforced by the result that the cusplike behaviour of $H_{c2}(\theta)$ becomes less pronounced well below $T_c$ (Fig. 3c), which is the opposite trend to $H_{c2}(\theta)$ behaviour of the conventional multilayer systems[21]. In fact Pauli-limited upper critical field $H_{c2}^{\mathrm{Pauli}}$ given by

$$H_{c2}^{\mathrm{Pauli}}=\sqrt{2}\Delta/g\mu_B \quad (1),$$

where $g$ is the gyromagnetic ratio determined by the Ce crystalline electric field levels, varies smoothly with field direction, consistent with $H_{c2}(\theta)$ of the present superlattices at low temperatures. Figure 4a displays the $H$-$T$ phase diagram of the superlattices. What is remarkable is that with decreasing $n$, $T_c$ decreases rapidly from the bulk value, while $H_{c2}$ does not exhibit such a reduction for both field directions. In fact, at low temperatures, $H_{c2\parallel}$ of $n$=5 and 7 is even larger than that of the bulk. This robustness of $H_{c2}$ (and hence of Δ) against $n$-reduction indicates that the superconducting pairing-interaction is hardly affected by two-dimensionalisation. This provides strong evidence that the superconductivity in bulk CeCoIn$_5$ is mainly mediated by 2D spin-fluctuations, although neutron spin resonance mode is observed at 3D (π,π,π) position below $T_c$[10].

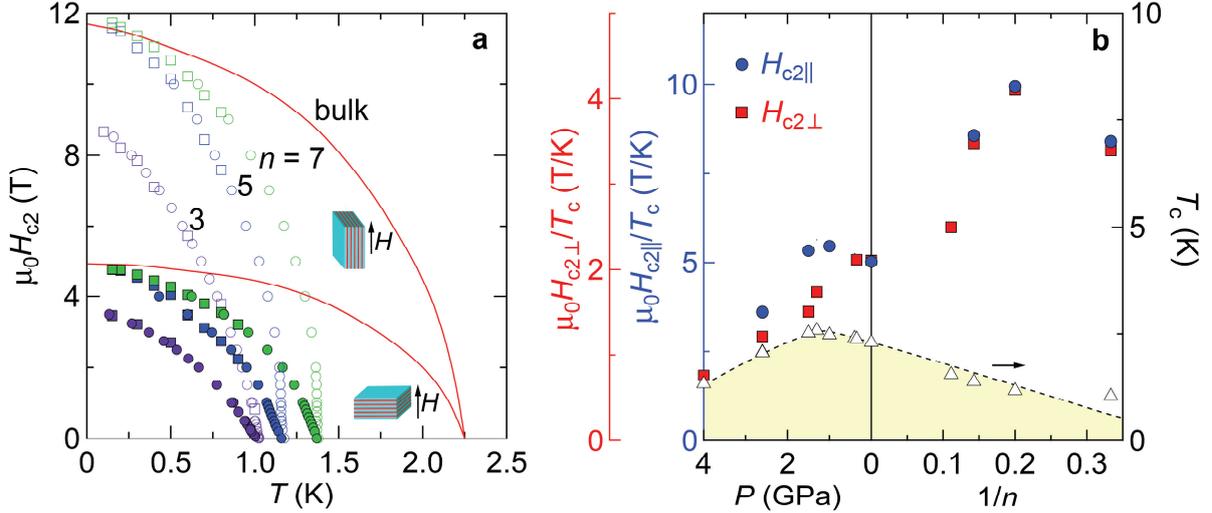

**Figure 4 | Superconducting phase diagrams of superlattices ($n$:5) of CeCoIn$_5$(n)/YbCoIn$_5$(5). a**, Magnetic field vs. temperature phase diagram of $n$=3, 5, and 7 superlattices in magnetic field parallel (open symbols) and perpendicular (closed symbols) to the *ab* plane compared with the bulk CeCoIn$_5$ data. The mid-point of the transition in the ρ($T$) (circles) and ρ($H$) (squares) has been used to evaluate $H_{c2}(T)$. **b**, Superconducting transition temperature, $T_c$ (open triangles), the reduced critical fields $H_{c2}/T_c$ in parallel (filled blue circles) and perpendicular (filled red squares) fields as a function of dimensionality parameter 1/$n$ (right panel). The pressure dependence of these quantities[28] is also shown for comparison (left panel).

In sharp contrast to $H_{c2}$, the thickness reduction dramatically enhances $H_{c2}/T_c$ from the bulk value (Fig. 4b). A comparison with the pressure dependence results[22], which represent the increased three dimensionality, reveals a $T_c$ dome and a general trend of enhanced $H_{c2}/T_c$ with reduced dimensionality. Through the relation of Eq. (1), this trend immediately implies a remarkable enhancement of Δ/$T_c$ by two-dimensionalisation. We note that the enhanced impurity scattering cannot be primary origins of the $T_c$ reduction, as these effects do not significantly enhance the Δ/$T_c$ ratio in $d$-wave superconductors[23]. This is supported by no discernible interdiffusion by TEM results and ρ$_0$ of superlattices in the same order as ρ$_0$ of the bulk CeCoIn$_5$. The reduction of $T_c$ may be caused by the reduction of density-of-states (DOS) in the superlattices, but this scenario is also unlikely because the DOS reduction usually reduces the pairing interaction, which results in the reduction of Δ.

Using the reported value of 2Δ/$k_BT_c$=6 in the bulk single crystal[9], 2Δ/$k_BT_c$ for the $n$=5 superlattice is estimated to exceed 10, which is significantly enhanced from the weak-coupling BCS value of 2Δ/$k_BT_c$=3.54. It has been suggested theoretically that $d$-wave pairing mediated by antiferromagnetic fluctuations in two-dimension can be much stronger than that in three-dimension[24-26]. The striking enhancement of 2Δ/$k_BT_c$ associated with the reduction of $T_c$, a situation resemblant to underdoped high-$T_c$ cuprates, implies that there appear to be additional mechanisms, such as 2D phase-fluctuations[27] and strong pair-breaking effect due to inelastic scattering[28]. Further investigation, particularly probing electronic and magnetic excitations in the normal and superconducting states, is likely to bridge the physics of highly unusual correlated electrons in the 2D Kondo lattice and in the 2D CuO$_2$ planes of cuprates. The fabrication in a wide variety of nanometric superlattices also opens up a possibility of nanomanipulation of heavy-electrons, providing a unique opportunity to produce a novel superconducting system and its interface.

**Method.** CeCoIn$_5$/YbCoIn$_5$ superlattices are grown by the molecular-beam-epitaxy (MBE) technique. The pressure of the MBE chamber was kept at 10$^{-7}$ Pa during the deposition. (001) surface of MgF$_2$ with rutile structure ($a$=0.462 nm, $c$=0.305 nm) was used as a substrate. The substrate temperature was kept at 550°C during the deposition. Atomic-layer-by-layer MBE provides for digital control of layer thickness, which we measure by counting the number of unit cells[2]. Each metal element was evaporated from individually controlled Knudsen-cells. 15-unit-cell-thick (uct) YbCoIn$_5$ was grown after CeIn$_3$ (28 nm) was grown on the (001) surface of substrate MgF$_2$ as a buffer layer. Then $n$-uct CeCoIn$_5$ layers and $m$-uct YbCoIn$_5$ (typically $m$=5) were grown alternately, typically repeated for 30-60 times. The deposition rate was monitored by a quartz oscillating monitor and the typical deposition rate was 0.01-0.02 nm/s.

**Acknowledgments.** We acknowledge discussions with R. Arita, A.V. Chubukov, M. J. Graf, P.A. Lee, N.P. Ong, S.A. Kivelson, T. Takimoto, and I. Vekhter. This work was supported by KAKENHI from JSPS and MEXT and by Grant-in-Aid for the Global COE program ``The Next Generation of Physics, Spun from Universality and Emergence'' from MEXT.

**SUPPLEMENTARY INFORMATION**

# Extremely Strong Coupling Superconductivity in Artificial Two-dimensional Kondo Lattices

**X-ray diffraction (XRD) analysis**

Figure S1 shows the XRD satellite peak near the (004) peak for (1:5) superlattice and the results of our simulation with various interdiffusion ratio. The intensity $I$ is calculated from the Laue function $G$ and crystal structure factor $F$:

$$I = |F|^2 G.$$

The crystal structure factor of the superlattice can be described as

$$F = F_{Ce} + F_{Yb},$$

where $F_{Ce}$ and $F_{Yb}$ are the crystal structure factors of CeCoIn$_5$ and YbCoIn$_5$, respectively. To estimate the Ce/Yb interdiffusion quantitatively, we assume that the interdiffuion occurs only at the CeCoIn$_5$/YbCoIn$_5$ interface layers (diffusion ratio $p$), ignoring the interdiffusion in the next layers. (The step model used in Fig. 1c corresponds to $p=0$.) Then $F_{Ce}$ and $F_{Yb}$ of the superlattice (1:5) having $l$ unit blocks perpendicular to the film plane are given by

$$F_{Ce} = (1-p)f_{Ce} + pf_{Yb} + f_{Co}\exp[\pi i q((1-p)c_{Ce} + pc_{Yb})]$$
$$+ f_{In} + 2f_{In}\exp(2\pi i q)[\exp((1-p)c_{Ce}z_{Ce} + pc_{Yb}z_{Yb}) + \exp((1-p)c_{Ce}(1-z_{Ce}) + pc_{Yb}(1-z_{Yb}))]$$

$$F_{Yb} = \left\{\exp[(1-p)f_{Ce} + pf_{Yb}] + \exp\left[2\pi i q\left(\left(1-\frac{p}{2}\right)c_{Ce} + \left(4+\frac{p}{2}\right)c_{Yb}\right)\right]\right\}$$
$$\times\left\{\frac{p}{2}f_{Ce} + \left(1-\frac{p}{2}\right)f_{Yb} + f_{Co}\exp\left[\pi i q\left(\frac{p}{2}c_{Ce} + \left(1-\frac{p}{2}\right)c_{Yb}\right)\right]\right.$$
$$\left. + f_{In} + 2f_{In}\exp(2\pi i q)\left[\exp\left(\frac{p}{2}c_{Ce}z_{Ce} + \left(1-\frac{p}{2}\right)c_{Yb}z_{Yb}\right) + \exp\left(\frac{p}{2}c_{Ce}(1-z_{Ce}) + \left(1-\frac{p}{2}\right)c_{Yb}(1-z_{Yb})\right)\right]\right\}$$

$$+ \sum_{j=1}^{3}\left\{\exp\left[2\pi i q\left(\left(1-\frac{p}{2}\right)c_{Ce} + \left(j+\frac{p}{2}\right)c_{Yb}\right)\right]\right\}$$
$$\times\{f_{Yb} + f_{Co}\exp(\pi i q c_{Yb}) + f_{In} + 2f_{In}\exp(\pi i q c_{Yb})[\exp(z_{Yb}) + \exp(1-z_{Yb})]\},$$

where

$$q = \frac{2\sin\theta}{\lambda}.$$

Here $\lambda$ is the wave length of the X-ray, $\theta$ is the diffraction angle, $f_{Ce}$, $f_{Yb}$, $f_{Co}$ and $f_{In}$ are the atomic form factors for Ce, Yb, Co and In, respectively, and $c_{Ce}$, $c_{Yb}$ are the $c$-axis lattice constants of CeCoIn$_5$ and YbCoIn$_5$, respectively. In the equation of $F_{Yb}$, 1$^{st}$, 2$^{nd}$ and 3$^{rd}$ lines correspond to the YbCoIn$_5$ layers at the interfaces (2$^{nd}$ and 6$^{th}$ layers of the (1:5) superlattice) involving Ce/La interdiffusion with a ratio $p/2$, 4$^{th}$ and 5$^{th}$ lines correspond to 3 pure YbCoIn$_5$ layers. The Laue function $G$ for the superlattice ($n:m$) having $l$ unit blocks is written as

$$G = \left\{ \frac{\sin[\pi q l(nc_{Ce} + mc_{Yb})]}{\sin[\pi q(nc_{Ce} + mc_{Yb})]} \right\}^2.$$

To compare with the XRD data, we further use the Gaussian distribution function for calculating $I(2\theta)$ to account for the instrumental resolution of 0.18 deg. The height of the satellite peaks normalized by the main (004) peak can be reproduced well by assuming no interdiffusion between Ce and Yb atoms near the interfaces ($p=0$). A 10 % interdiffusion makes the peak height discernibly lower.

Combining the TEM, and XRD analyses, we conclude that the superlattice structures are realized in an epitaxial form without noticeable interdiffusions.

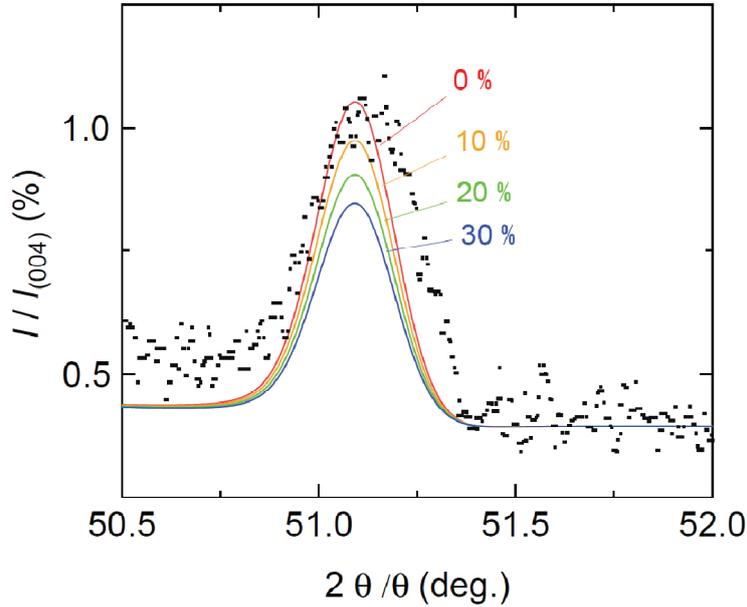

**Supplementary Figure S1.** XRD satellite peaks (black dots) around (004) main peak in the (1:5) superlattice with 60 unit blocks. The satellite peaks are compared with our simulations assuming that 0 (red solid line), 10 (orange), 20 (green) and 30 % (blue) of Ce atoms at the interfaces are replaced with Yb. We used $c$-axis lattice constants of 0.75513 nm for CeCoIn$_5$ and 0.7433 nm for YbCoIn$_5$.

**Electronic band structure of the superlattice**

We investigate the electronic band structure in the (1:5) superlattice based on the *ab initio* density functional theory. The band structure calculations for $Ln$CoIn$_5$ ($Ln$=Ce, La, and Yb) and for the superlattice have been performed by using the relativistic full-potential (linearized) augmented plane-wave (FLAPW) + local orbitals method as implemented in the WIEN2k package[S1]. The crystallographical parameters used in the calculations are summarized in Table S1.

Figures S2(a)-(h) display the energy dispersion, $E_k$, along the high-symmetry line. In the Yb case, the almost dispersion-less bands between 0 and -2 eV come from the 4$f$ electrons, which do not cross the Fermi level (Fig. S2(c)). This indicates that the 4$f$ electrons do not participate in the conduction, which is consistent with the absence of Kondo behavior in the transport properties in YbCoIn$_5$. In contrast, the 4$f$ bands extend to the Fermi level in the Ce case (Fig. S2(a)), in which the Kondo effect is clearly observed. This difference between Yb and Ce is reinforced by the calculations of partial

charges in the atomic spheres (Table S2). In the case of Ce, the number of 4$f$ electrons inside the Muffin-tin sphere is estimated as 0.95 in the FLAPW calculations. This number is very close to exact unity, as we expect some protrusion (~5%) outside the sphere. In YbCoIn$_5$, we should also consider that the obtained number of 13.45 for $f$ electrons is actually close to 14, which is in the closed-shell configuration. (In fact we verified this fact by an additional calculation of the 4$f$ open-core treatment.) In the superlattice, we found that these occupation numbers of $f$ orbitals in each layer are essentially unchanged from the parent materials. This indicates that mobile 4$f$ electrons are confined only in the Ce layers.

Figures S3(a)-(d) depict the partial density of states (DOS), whose values at the Fermi level are reported in Table S2. The DOS at the Fermi level in the superlattice is nearly equal to the total of the DOS in the parent materials, $26.83 + 8.70 \times 5 = 70.33$ (mJ/mol K$^2$). In addition, the partial DOS of Ce component is also comparable to that in CeCoIn$_5$.

In Figs. S4(a)-(d), we show the Fermi surface colored by the Fermi velocity, $v_{kF}=|(v^x_{kF},v^y_{kF},v^z_{kF})|$, where $v^i_{kF} = 1/\hbar(dE_{kF}/dk^i_F)$. The blue parts of Fermi surface have smaller Fermi velocity, and then larger weight of Ce(4$f$) component. The Fermi surface of the superlattice (Fig. S4(d)) is much more two-dimensional than the 3D structure in the parent materials, which is mostly due to the folding along the $c$-axis. We note that the cylindrical bands near the zone center ($\Gamma$ point) are hole-like Fermi surface whilst the bands near the zone corner (M point) are electron-like (see the band dispersions in Fig. S2(h)). This shape of these sheets suggests a good nesting between these bands, which would enhance the susceptibility at a Q-vector of ($\pi, \pi, 0$). It is tempting to suggest that such 2D antiferromagnetic fluctuations are responsible for the observed strong-coupling 2D superconductivity in the superlattices, which deserves further studies.

**Supplementary Table S1** | All materials studied here have the space group No.123 P4/mmm. Lattice constants in the superlattice CeCoIn$_5$(1)/YbCoIn$_5$(5) are determined by $a=(a_{CeCoIn5}+5a_{YbCoIn5})/6$ and $c=c_{CeCoIn5}+5c_{YbCoIn5}$, and we keep the relative atomic positions (Wyckoff positions) in each layer the same as those in the parent materials.

|  | $a$ (Å) | $c$ (Å) | Wyckoff position $z$ of In(2) |
|---|---|---|---|
| CeCoIn$_5$ [S2] | 4.612 | 7.549 | 0.305 |
| LaCoIn$_5$ [S3] | 4.634 | 7.615 | 0.311 |
| YbCoIn$_5$ [S4] | 4.559 | 7.433 | 0.305 |
| (1:5) superlattice | 4.568 | 44.714 | --- |

**Supplementary Table S2** | Occupation number of $d$ and $f$ orbitals inside the Muffin-tin spheres of $Ln$ and Co atoms and the DOS at the Fermi level. Occupation number in the superlattice is described as Ce/Yb. Five Yb atoms have almost the same occupation number. In the last column, values in the parentheses represent the partial DOS of Ce components.

|  | $Ln(f)$ | $Ln(d)$ | Co($d$) | DOS(mJ/mol K$^2$) |
|---|---|---|---|---|
| CeCoIn$_5$ | 0.950 | 0.718 | 7.54 | 26.83 (14.5) |
| LaCoIn$_5$ | 0.081 | 0.646 | 7.55 | 6.81 |
| YbCoIn$_5$ | 13.45 | 0.456 | 7.58 | 8.70 |
| (1:5) superlattice | 0.986 / 13.48 | 0.712 / 0.456 | 7.59 | 69.36 (13.4) |

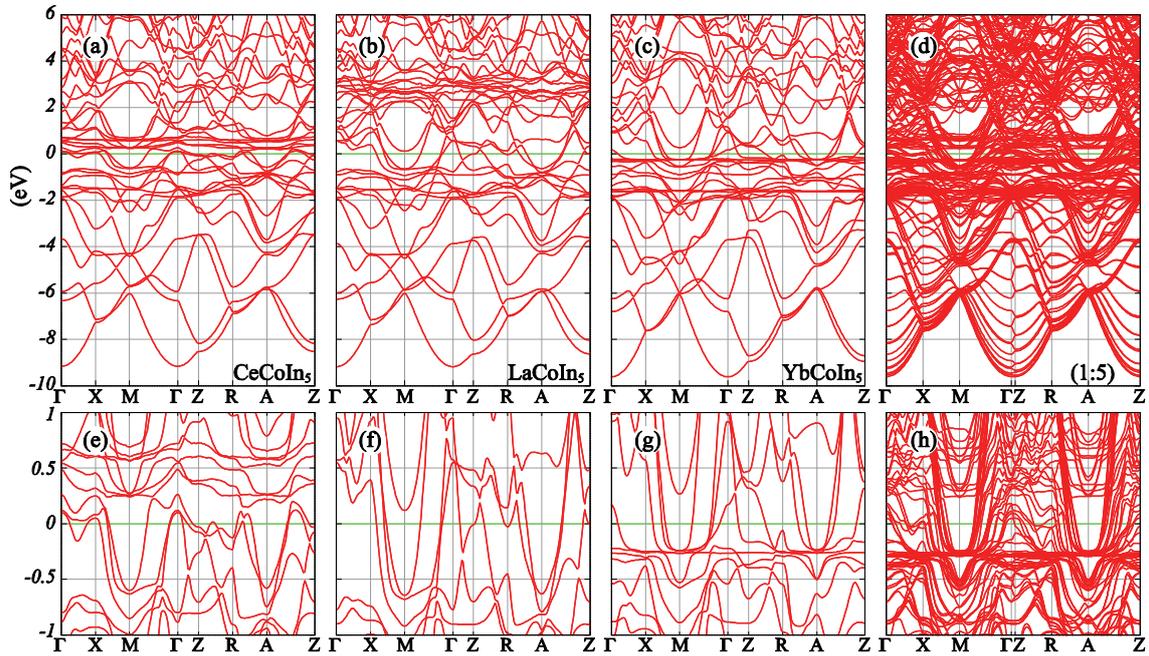

**Supplementary Figure S2** | Energy band dispersion along the high-symmetry line in CeCoIn$_5$ (a), LaCoIn$_5$ (b), YbCoIn$_5$ (c), and the (1:5) superlattice (d). The Fermi level is indicated by the green line. Lower panels (e-h) are the enlarged figures near the Fermi level.

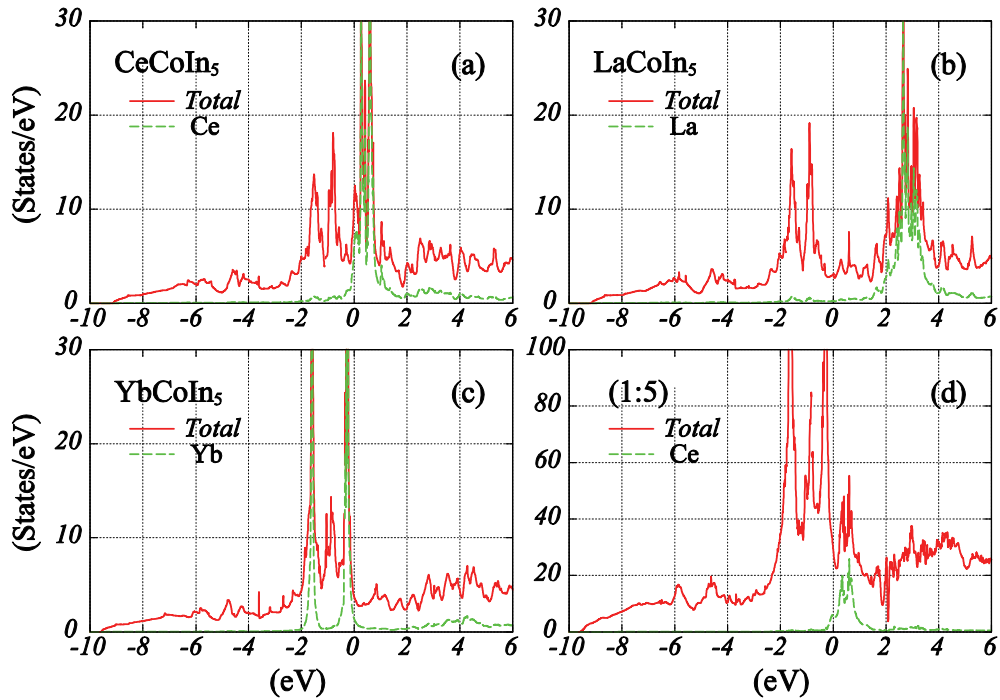

**Supplementary Figure S3** | DOS in the unit of eV. The partial DOS of *Ln* ions is indicated by green dashed lines.

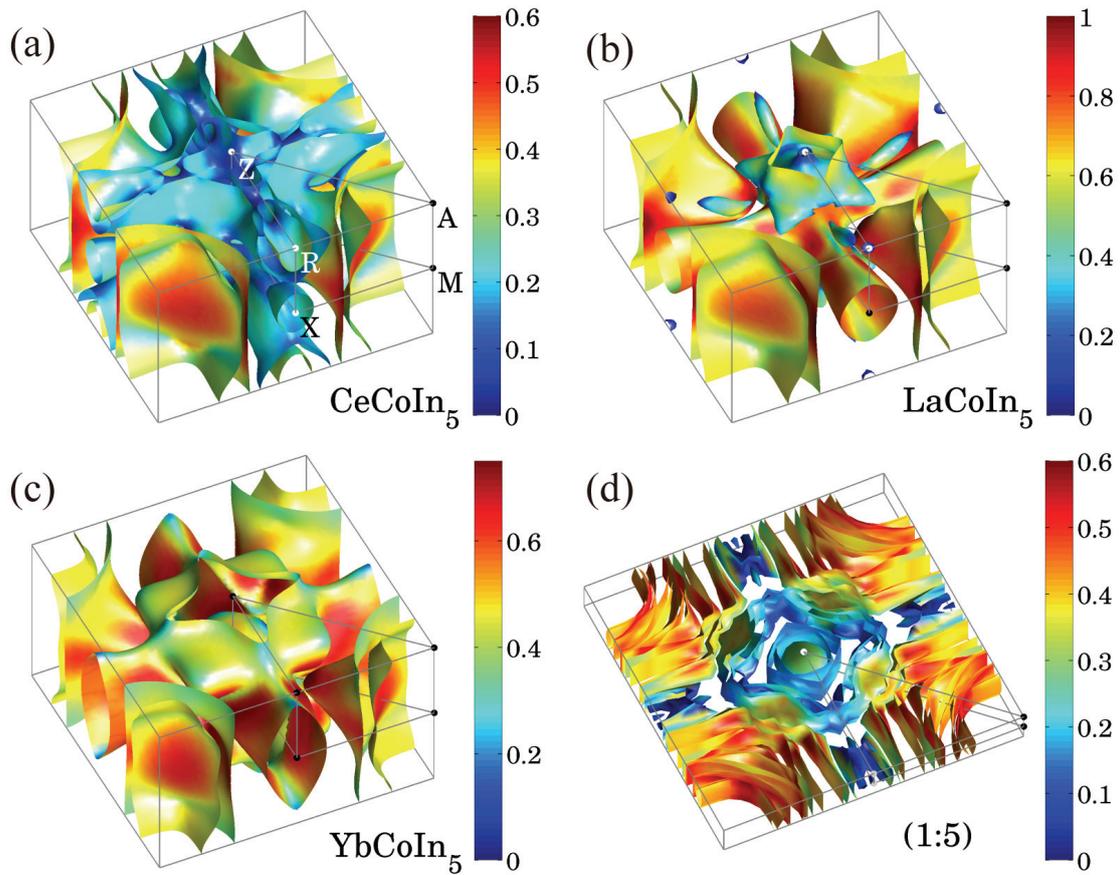

**Supplementary Figure S4** | The Fermi surfaces colored by magnitude of the Fermi velocity. The unit is $10^6$ (m/sec).

**SUPPLEMENTARY REFERENCES**
S1. Blaha, P., Schwarz, K., Madsen, G. K. H., Kvasnicka, D. & Luitz, J., WIEN2k, An Augmented Phane Wave + Local Orbitals Program for Calculating Crystal Properties (Karlheinz Schwarz, Techn. Universitat Wien, Austria, 2001).
S2. Settai, R., *J. Phys.: Condens. Matter* **13**, L627 (2001).
S3. Macaluso, R. T., *J. Solid State Chem.* **166**, 245 (2002).
S4. Zaremba, V. I., *Z Anorg. Allg. Chem.* **629**, 1157 (2003).